\documentclass{iau-JDSS}
\usepackage{graphicx}
\usepackage{url}
\def\aap{{\it A\&A}}
\def\aj{{\it AJ}}
\def\apj{{\it ApJ}}
\def\apjl{{\it ApJL}}

\def\mnras{{\it MNRAS}}
\def\Fermi{\textit{Fermi}}
\def\Planck{\textit{Planck}}

\title[Radio \& High Energy Connection in AGNs] 
{JD6 - The Connection between Radio Properties and High Energy Emission in AGNs}

\author[G.~Giovannini \& T.~Cheung]   
{Gabriele Giovannini$^{1,2}$
\and Teddy Cheung$^3$}

\affiliation{$^1$Department of Physics and Astronomy, Bologna University, Italy
\break email: ggiovann@ira.inaf.it\\[\affilskip]
$^2$ Istituto di Radioastronomia-INAF, Italy\\[\affilskip]
$^3$ Naval Research Laboratory, Washington, DC 20375, USA
\break email: Teddy.Cheung@nrl.navy.mil}

\pubyear{2012}
\volume{Volume 16}  
\pagerange{.....}
\date{?? and in revised form ??}
\jname{Highlights of Astronomy, Volume 16}
\editors{Ian F. Corbett, ed.}
\begin{document}

\maketitle

\begin{abstract}
While observations in the radio band are providing essential information on the 
innermost structures of relativistic jets in active galactic nuclei (AGN), the 
recent detection by \Fermi\ of gamma-ray emission from many hundreds of blazars 
shows that the maximum jet power is emitted at high energies. Multi-wavelength 
monitoring observations further allow variability studies of the AGN spectral 
energy distributions over 13 orders of magnitude in frequency. The Joint 
Discussion offered the possibility for a comprehensive discussion of advances in 
the observational domain and stimulated theoretical discussion about our current 
understanding of jet physics.
\end{abstract}

\firstsection 
\section{Introduction}

A Joint Discussion (JD) dedicated to the connection between radio properties and 
high energy emission in active galactic nuclei (AGN) took place on the 23rd and 
24th of August 2012 during the XXVIII IAU General Assembly in Beijing, China. It 
was a great opportunity to confront observational multiwavelength data with 
theoretical models for AGN. The JD included four main sessions covering the AGN 
population in the radio and gamma-ray bands (\S~2), high resolution core and jet 
properties (\S~3), multiwavelength correlations and variability (\S~4), and a 
final discussion on jet physics and the role of black hole (BH) spin and 
accretion (\S~5). Eight invited plus 11 contributed oral talks were presented 
together with 34 contributed posters to an interesting scientific discussion. In 
the following we present a brief summary for the talks in the four sessions.

All presentations are available at: 
\url{http://www.ira.inaf.it/meetings/iau2012jd6/Program.html} and poster 
abstracts in: \url{http://www.ira.inaf.it/meetings/iau2012jd6/Posters.html}.

\section{Session 1: The AGN population as seen in the Radio and Gamma-ray bands}

\subsection{T. Cheung -- The AGN population in Radio and Gamma-rays: Origins and 
Present Perspective (invited)}

\begin{figure}
\begin{center}
\includegraphics[width=4in]{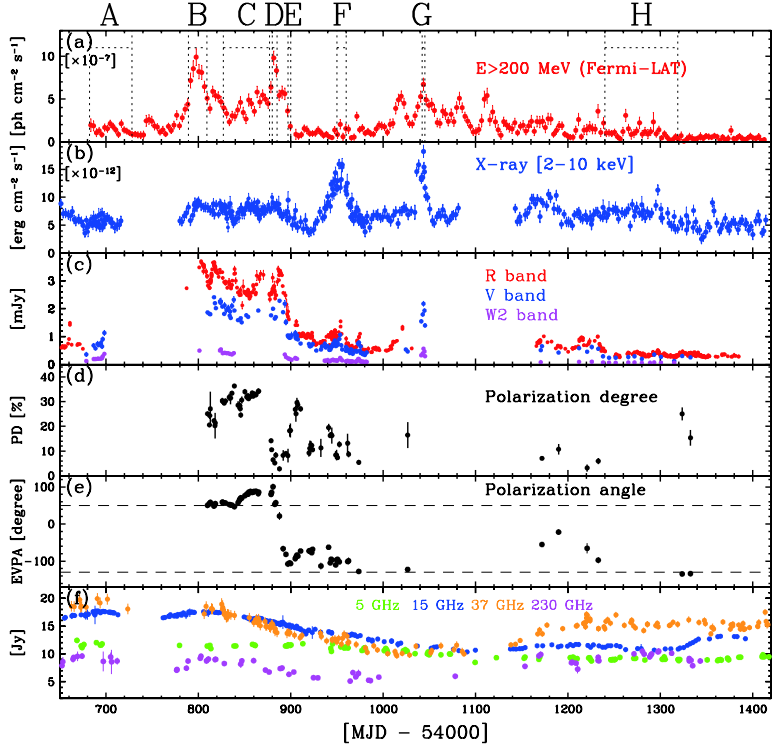}
  \caption{Multi-band (gamma-ray, X-ray, optical, including polarization degree
and EVPA, and radio lightcurves of 3C~279 from 2008 August to 2010 August. 
Taken from \cite{hay12}.}
    \label{fig:hayashida}
\end{center}
\end{figure}

Radio observations present a rich phenomenology in studies of AGN -- extensive 
multi-frequency lightcurves (e.g., Fig~\ref{fig:hayashida}), observations and 
statistics of superluminal motions, radio spectral polarimetric imaging on all 
scales (jets, hotspots, lobes), and an abundant variety of source types 
including young radio sources and radio-quiet objects. In the EGRET era, there 
were 66 high-confidence blazars identified with gamma-ray sources (27 
lower-confidence) and only a few radio galaxies (e.g., Cen A, 3C~111). Now we 
are in the era of the \Fermi\ Gamma-ray Space Telescope with the increased 
capabilities provided by the Large Area Telescope (LAT). The 2nd LAC AGN catalog 
(2LAC) ``clean'' sample included: 310 flat spectrum radio quasars (FSRQs), 395 
BL Lacs, 156 blazars of unknown type, and 24 AGNs (include radio galaxies). 
FSRQs have been detected with $L_\gamma$ $\sim$10$^{48}$ erg/s up to $z \sim 3$. 
Significant correlations are present between radio and gamma-ray properties. 
FSRQs are on average brighter and apparently more luminous in the radio band 
than BL Lacs (but a redshift incompleteness is present).

Thanks to the 1st \Fermi-LAT Hard source Catalog (1FHL) which includes data from 
August 2008 through July 2011, possibly softer AGN gamma-ray spectra appear with 
increasing redshift, and many new targets are now available for current and 
future TeV telescopes. The high energy emission site for AGN can be probed with 
gamma-ray imaging in exceptional cases, like in the giant radio lobes of 
Centaurus A (and possibly in NGC 6251, and Fornax A). Moreover \Fermi\ detected 
giant gamma-ray bubbles in our Galaxy (\S~\ref{sec:su}) and a possible young 
radio source (4C+55.17). In the coming years, it will be possible to extend the 
radio/gamma-ray correlations to low fluxes/luminosities, to continue to identify 
possible sites of gamma-ray emission, to test if radio-quiet AGN are also 
gamma-ray quiet, and to look for bubble sources in nearby galaxies.

References: \cite{abd10a}, \cite{ack11b}, \cite{mcc11}, \cite{pan12}, 
\cite{tak12}.

\subsection{\L. Stawarz -- The AGN population in Radio and Gamma rays: 
theoretical perspectives (invited)}

Radio-loud systems appear to be the only AGN loud in gamma-rays. In addition to 
blazars and radio galaxies, the only newly established class of gamma-loud AGN 
are the radio-loud narrow line Seyfert 1s (NLS1s). Radio-quiet Seyferts seem 
gamma-quiet as well. Radio properties of nuclear jets may be directly related to 
gamma-ray properties. However, nuclear relativistic jets are not the only 
relevant gamma-ray production site as lobes and bubbles are observed in 
gamma-rays. Correlations between radio and gamma-ray properties seem to be 
present, but more data and better statistics are necessary to understand if they 
are real or not. In flux-flux limited samples, artificial correlations are 
expected. Evidence of co-spatiality is not firmly established.

References: \cite{abd09b}, \cite{ack12}, \cite{ghi11}, \cite{hess12}

\subsection{H. Sol -- Very high energy gamma-ray radiogalaxies and blazars 
(contribution)}

At present, 53 firmly known very high energy (VHE; $>0.1$ TeV) AGN have been 
observed. Among them we have 47 blazars, 4 radio galaxies and 1 AGN of unknown 
type and possibly Sgr A*. The blazar sample includes: 34 high-frequency peaked 
BL Lacs (HBL), 4 intermediate-frequency peaked BL Lacs (IBL), 4 low-frequency 
peaked BL Lacs (LBL), 3 FSRQ, and 2 BL Lacs. The four radio galaxies are: M87, 
Cen A, NGC 1275, IC 130. Most of them are beamed sources with a strong Doppler 
boosting as expected since it helps to accommodate fast variability and to avoid 
strong intrinsic absorption. Variability time scales are from a few minutes to 
months and years. Multi-zone synchrotron self-Compton (SSC) models can reproduce 
most of HBL stationary state spectra. 

References: \cite{acc11}, \cite{ale12}, \cite{abr12}.

\subsection{D. McConnell -- Counterparts to \Fermi-LAT sources from the ATPMN 5 
and 8GHz catalogue of southern radio sources (contribution)}

The Australia Telescope Parkes-MIT-NRAO (ATPMN) catalogue gives accurate 
positions and flux density measurements at 4.8 and 8.6 GHz for 8385 sources, 
with a typical image rms noise of 2 mJy/beam. Significant long term variability 
for 85$\%$ of the sources was detected using the 10-14 year baseline between the 
ATPMN and AT20G observations. Short term variations were gauged by inspecting 
ATPMN scans separated by up to 4-8 hrs, with about 30$\%$ showing intra-day 
variability. Polarization is available for 9040 sources. Among them $\sim$126 
sources have been identified with 2FGL sources and 112 have an optical 
identification within 3$''$ from the UK Schmidt survey.

Reference: \cite{mcc12}.

\begin{figure}
\begin{center}
\includegraphics[width=4.2in]{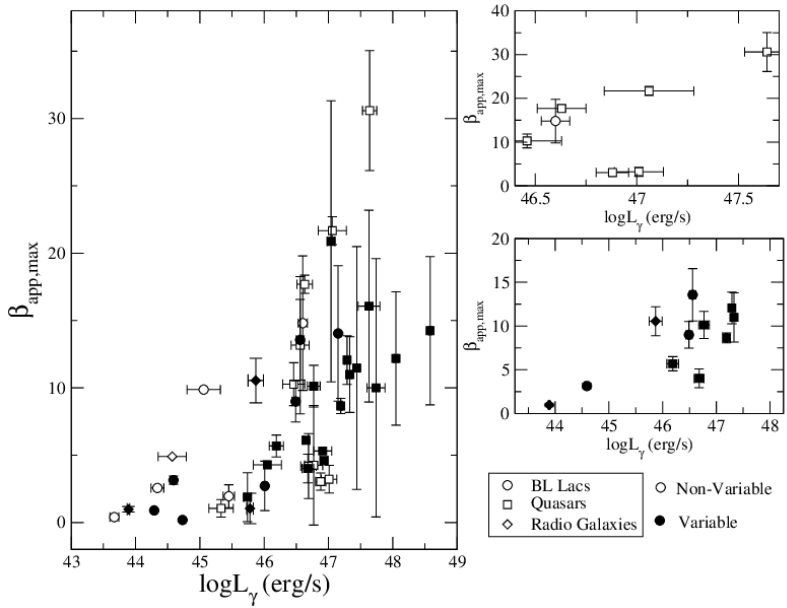}
  \caption{Maximum VLBI measured apparent velocities for gamma-ray detected 
sources from the CJF sample versus gamma-ray luminosity (filled symbols 
are gamma-ray variable while open symbols are non-variable in gamma-rays). 
Taken from \cite{kar11}.}
    \label{fig:karouzos}
\end{center}
\end{figure}

\subsection{M. Karouzos -- Gamma-rays in flat-spectrum AGN: revisiting the fast 
jet hypothesis (contribution)}

Based on studies using EGRET and the early \Fermi-LAT AGN samples, gamma-ray 
detected AGN were found to show on average faster apparent speeds with respect 
to other AGN not detected in gamma-rays. To confirm and investigate this, the 
very long baseline interferometry (VLBI) proper motion results for 198 QSOs and 
33 BL Lacs from the Caltech Jodrell Bank Flat-spectrum (CJF) 5 GHz VLBI survey 
were analyzed (Fig.~\ref{fig:karouzos}). Among these 61 sources have been 
detected by \Fermi-LAT, consisting of 32 FSRQs, 24 BL Lacs, and 5 radio 
galaxies. Conclusions are: i) no strong link is present between fast jets and 
gamma-ray detection; ii) AGN class and gamma-ray variability are connected to 
jet speeds; iii) a correlation between gamma-ray luminosity and apparent 
velocity is found (higher velocity for stronger gamma-variable sources); iv) 
gamma-ray detected sources appear wider and with larger jet distortions. 
Different findings with respect to the previous studies may be due to different 
observing frequency (probing either different jet regions or structures) or the 
difference in sampling of the proper motion data.

Reference: \cite{kar11}.

\subsection{F. D'Ammando -- To be or not to be a blazar. The case of gamma-ray 
narrow line Seyfert 1 SBS~0846+513 (contribution)}

In 2008, the first NLS1 PMN J0948+0022 was detected by the \Fermi-LAT. After 
that another four NLS1s were detected in gamma-rays. These results confirm the 
presence of relativistic jets also in NLS1s even though these sources are 
typically thought to be hosted in spiral galaxies. Their average 
spectral energy 
distributions (SEDs) are similar to FSRQs, but at lower luminosity. 
SBS 0846+513 is a new gamma-ray NLS1 clearly 
detected during the third year of \Fermi\ operation, 
in particular during a flaring state in 2011
June-July. The 
gamma-ray peak on daily timescale corresponds to an isotropic luminosity
of about 10 $^{48}$ erg s$^{-1}$, 
comparable to that of luminous FSRQs. While the kpc-scale structure is 
unresolved in VLA images, there is a core-jet structure seen in VLBA images. The 
mechanism at work for producing a relativistic jet in NLS1s is not clear. 
Fundamental parameters should be the BH mass and the BH spin. This source could 
be a blazar with a BH mass at the low end of the blazar's BH mass distribution. 
Gamma-ray NLS1s have larger masses with respect to the entire sample of NLS1s. 
Moreover prolonged accretion episodes could spin-up the SMBH leading to a 
relativistic jet formation.

References: \cite{abd09a}, \cite{dam12}.

\begin{figure}
\begin{center}
\includegraphics[width=3in]{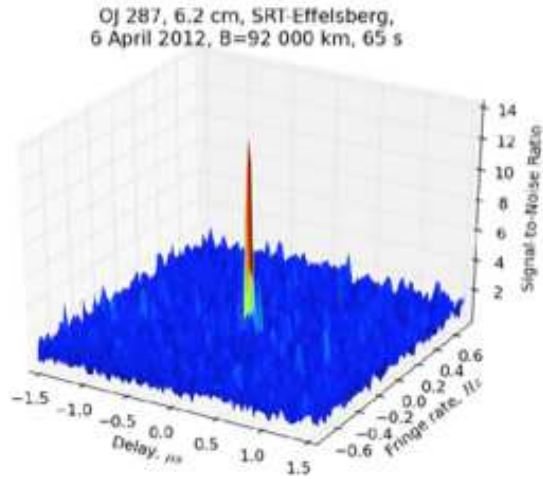}
  \caption{Early fringe obtained for OJ~287 at 6.2cm between the RadioAstron 
space radio telescope and the ground-based Effelsberg 100m with a baseline of 
7.2 Earth diameters. Taken from \cite{kar13}.}
    \label{fig:kardashev}
\end{center}
\end{figure}

\section{Session 2: High resolution core and jet properties}

\subsection{N. Kardashev -- RadioAstron Space VLBI mission: early results 
(invited)}

RadioAstron was launched on 18th July 2011. The orbital period is 8.5 days, 
with a perigee radius = 67,000 km and an apogee = 282,000 km. Observations are 
carried out at P, L, C, and K bands in dual-polarization, with finest angular 
resolutions (300,000 km baseline) of 580, 113, 39, and 7.5-10 $\mu$-arcsec, 
respectively. Among the first sources detected was a single giant pulse in PSR 
B0950+08 at 92 cm, with significant variations in 1 hr due to ISM plasma 
scintillations. The first fringes in K-band were observed together with 
baselines to the Effelsberg 100m from the quasar 2013+370; see 
Fig.~\ref{fig:kardashev} for another example. Imaging of the quasar 0716+714 at 
4.8 GHz was obtained using the space radio telescope and eight ground telescopes 
showing the core-jet structure as seen in previous VLBI images. A status and 
more recent summary of the RadioAstron results appeared recently.

Reference: \cite{kar13}.

\subsection{M. Lister -- Blazars at high resolution: what large multi-epoch 
VLBI studies can tell us (invited)}

Blazar studies suffer heavy sample selection effects: e.g., obscuration in 
optical and X-ray; spectral contamination from accretion disk emission and from 
lobe (unbeamed) emission; non simultaneous observations; and more. To address 
blazar sample biases it is important to concentrate on `uncontaminated' bands. 
The \Fermi\ 2LAC AGN catalog has no contamination from the host galaxy, even if 
may still be incomplete due to issues with source associations. The MOJAVE VLBA 
program provides regular observations of radio-bright AGNs at 15 GHz. With 24 
hrs observing sessions every 3 weeks, it assures continuous time baseline data 
on many sources back to 1994. Among the main results we quote: the brightest 
gamma-ray and radio-selected quasars have similar redshift distributions; 
gamma-ray selected blazars have an additional sub-population of low-redshift HSP 
BL Lacs that are intrinsically very bright in gamma-rays; lowest luminosity BL 
Lacs (HSPs) all have high gamma-ray loudness (due to SED peak location). In BL 
Lacs (HSP and LSP) the photon index is well correlated with the Compton peak 
location. This trend could not exist if the gamma-ray and pc scale radio jet 
emission were fully independent. Analyzing kinematics of 889 discrete features 
in 201 jets from 1994 to 2011, we derive that: jets of HSP BL Lacs are 
characterized by lack of compact superluminal features; BL Lac jets have lower 
radio synchrotron luminosity and lower speeds.

References: \cite{lis11}, \cite{ack11b}.

\subsection{M. Orienti -- On the connection between radio and gamma rays. The 
extraordinary case of the flaring blazar PKS 1510-089 (contribution)}

The FSRQ, PKS 1510-089 ($z=0.361$), shows strong variability and highly 
superluminal jet components found close in time with gamma-ray flares. Moreover, 
it was detected at VHE gamma-rays, shows a high level of polarized emission, and 
a large rotation of the electric vector position angle (EVPA) close in time with 
a gamma-ray flare. PKS 1510-089 underwent a very active period in 2011 reaching 
its historical maximum flux density in October 2011. The gamma-ray flare in July 
2011 occurred after a rotation of 380 degree of the optical EVPA suggesting a 
common region for the optical and gamma-ray emission. The new jet component is 
likely evidence of a shock propagating downstream along the jet. If the 
gamma-ray flare in October 2011 is related to the radio outburst, it would 
strongly support the idea that some gamma-ray flares are produced parsecs aways 
from the nucleus. Note that not all flares have the same characteristics, 
suggesting shocks with different properties. Follow-up in the mm regime with a 
high sensitivity VLBI array including ALMA will be crucial in determining the 
high-energy emitting region.

Reference: \cite{ori13}.

\subsection{Z. Abraham -- The radio counterparts of the 2009 exceptional 
gamma-ray flares in 3C~273 (contribution)}

A very strong and complex gamma-ray flare was observed in 3C~273 in September 
2009. The flare was related to the formation of superluminal components studied 
at 43 GHz and to a radio flare observed with the Itapetinga radio telescope at 
43 GHz. The gamma-ray flux increased by a factor of 20, while only by a factor 
two in radio. We explain this fact as a change in the Doppler factor, due to a 
change in the angle between the jet and the line of sight. This fact was 
predicted by the precessing jet model of Abraham \& Romero (1999). The 
precessing model based on the jet curvature and proper motions is compatible 
with a period of 16 years.

Reference: \cite{abr99}, \cite{jor11}.

\subsection{Z. Shen -- High precision position measurements of the cores in 
3C~66A and 3C~66B (invited)}

3C~66A is a low frequency peaked BL Lac object at $z=0.444$. It is characterized 
by prominent variability at radio, IR, and optical frequencies. It shows a 
one-sided core-jet structure with detected superluminal motion. The core shift 
with frequency has been estimated in 2001 and 2006. A large difference has been 
found between the two measurements possibly due to a strong flux density 
increase at 15 GHz in 2006. This radio flare is possibly due to the core 
activity as shown by a new component that emerged from the central core region. 
Because of their small angular distance in the plane of the sky, 3C~66A and the 
radio galaxy 3C~66B, are an ideal pair to obtain a combined core-shift 
measurement of the two sources. Comparison with data at different epochs is 
confusing: the difference in core shift result cannot be simply explained by the 
core flare activity. Other parameters apart from core flux variability may 
influence the core shift. More data are required to explore this point.

References: \cite{cai07}, \cite{sud11}, \cite{zha11,zha13}.

\section{Session 3: Multi-wavelength correlations and variability}

\subsection{F. Tavecchio -- Variability of blazars: probing emission regions and 
acceleration processes (invited)}

Blazars show variability on timescale of days suggesting a parsec scale emission 
region. Comparing the radio-gamma variability, there is evidence that this 
active region should be inside the broad line region (BLR). Moreover, in some 
sources the GeV emission shows spectral breaks that could be due to absorption 
effects inside the BLR. In this context, a relevant case to consider is that of 
PKS 1222+216 (4C+21.35) which doubled its TeV flux on a timescale on the order 
of 10 minutes (Aleksi{\'c} et al. 2011). In this source the location of the VHE 
emission inside the BLR is problematic because the too strong absorption 
expected (e.g., Liu et al. 2008, Reimer 2007, Tavecchio \& Mazin 2009, Poutanen 
\& Stern 2010) due to the huge optical depth of the BLR. Possibilities to 
reconcile rapid variability in regions at large distances from the core (outside 
the BLR) require the presence of jet substructure (e.g., Ghisellini \& Tavecchio 
2008). Many models have been proposed but not all problems are solved -- e.g., 
the presence of mini-jets from fast reconnection in a highly magnetized jet 
(Giannios et al. 2009, 2010), or narrow electron beams from magnetocentrifugal 
acceleration (e.g., Ghisellini et al. 2009), beams from relativistic 
reconnection (e.g., Cerutti et al. 2012, Nalewajko et al. 2012), or ultra-high 
energy (UHE) neutral beams (Dermer et al. 2012). In conclusion, rapid 
variability is perhaps currently the most compelling issue in high-energy 
astrophysics. The idea of a unique, large, ``relaxed'' emission region is, at 
least sometimes, inadequate.

\begin{figure}
\begin{center}
\includegraphics[angle=-90,width=4.5in]{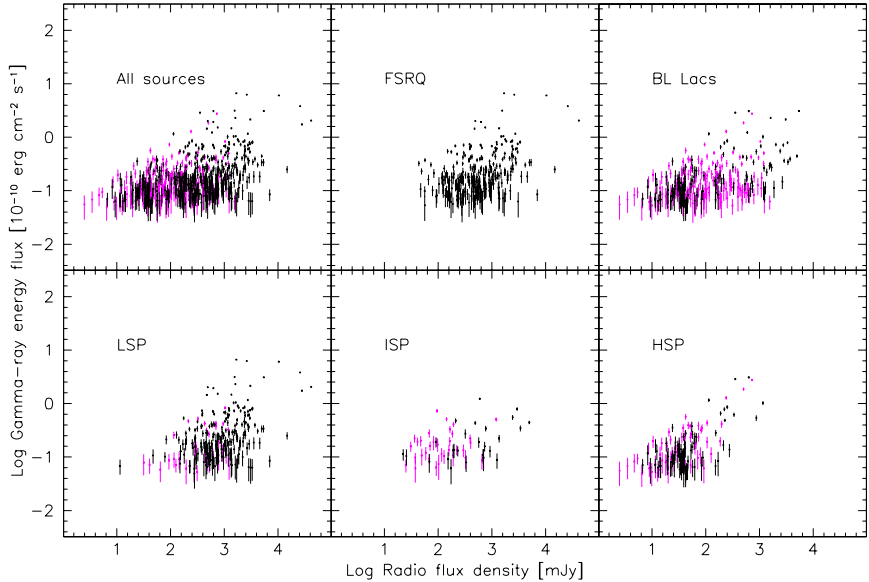}
\caption{Broad band gamma-ray energy flux vs.\ 8 GHz archival radio flux 
densities for the 1LAC sample separated by source type. Taken from 
\cite{ack11a}.}\label{fig:lat2011}
\end{center}
\end{figure}

\subsection{T. Hovatta -- Assessing AGN Variability and Cross-waveband 
Correlations in the Era of High-Quality Monitoring Data in Low and High Energies 
(invited)}

A relevant problem in AGN is where the high-energy emission is located: close to 
the BH within the BLR or further down in the parsec-scale jet? Correlations can 
be used to locate the unresolved gamma-ray emission site. We can have a 
flux-flux correlation (amplitude domain; e.g., Fig.~\ref{fig:lat2011}) or a 
light curve cross-correlation (time domain). Using simultaneous data, an 
intrinsic radio/gamma-ray flux density correlation is confirmed. FSRQs and BL 
Lacs show a different behavior (a possible selection effect?). Using archival 
non-contemporaneous data to increase the statistics, the correlation persists 
(it is even stronger for BL Lacs because of more sources).

More difficult is to tell if individual events are correlated and what are the 
time delays. Light curve correlations are difficult to establish in single 
sources. Opacity effects are important in the radio bands, moreover, we could 
still have too short time series. Stacked correlations show statistically 
significant time delays with increasing delays for longer wavelengths. Good 
multiwavelength coverage is really necessary to address this issue.

References: \cite{kov09}, \cite{ack11a}, \cite{mah10}, \cite{nie11}, 
\cite{max12}, \cite{pav12}.

\subsection{X. Liu -- VLBI core flux density and position angle analysis of the 
MOJAVE AGN (contribution)}

The two-fluid jet model assumes that:
1)  the outflow consists of an electron−proton plasma (the jet), moving at mildly
relativistic speed,
2) an electron-positron plasma (the beam) is moving at highly relativistic
speed, and 
3) the magnetic field lines are parallel to the flow in the beam and the
mixing layer, and are toroidal in the jet.

To confirm and investigate this jet structure we model-fit the MOJAVE blazar 
core sample which includes blazars with more than 10 years of VLBA monitoring, 
and more than 15 observed epochs with a good time distribution. This sample 
consists of 104 sources, 77 of which are quasars, 27 BL Lacs, in which 82 are 
\Fermi-LAT detected, and 22 non-detected sources. Of these, nine are also TeV 
sources. The model-fit result of the cores of 104 blazars from the MOJAVE 
monitoring data, suggests that \Fermi\ LAT-detected blazars have wider position 
angle changes of the inner-jet than LAT non-detected blazars, and are 
preferentially associated with higher variable blazars.

A two-zone jet model can explain the correlations in the model-fitted 
parameters. The \Fermi\ GeV gamma-ray detection rate show equally similar 
fraction for sources dominated by the innermost jet (zone-1) and sources 
dominated by the outer jet (zone-2). But importantly, TeV gamma-ray sources are 
associated mostly with blazars dominated by the outer part of inner-jet 
(zone-2).

Reference: \cite{lis09}.

\subsection{E. Valtaoja -- Gamma-ray emission along the radio jet: studies with 
Planck, Mets{\" a}hovi and \Fermi\ data (contribution)}

If the region emitting at VHE is close to the black hole - accretion disk region 
(i.e., it is inside the BLR), we expect that gamma-ray flares precede radio 
variations (assuming as VLBI zero epoch the beginning of a millimeter flare), 
and little or no correlation with radio variations. If the VHE region is 
distant, at or downstream of the radio core (i.e. outside the BLR), we expect 
gamma-ray flares simultaneous, or after, the beginning of radio variations and a 
correlation between VHE and radio variations.

Comparing data from \Fermi\ and the Mets{\" a}hovi radio sample, the case for 
``distant'' gamma-ray origin appears much stronger. Direct observational 
evidence for ``close'' origin was not found; observations point towards distant 
origins: in the radio-gamma correlations, there are evident delays from radio to 
gamma. A confirmation of this result will be possible from the final \Planck\ data 
to model the radio to gamma-ray SEDs with unprecedented accuracy.

Reference: \cite{leo11}.

\section{Session 4: Jet physics and the role of BH spin and BH accretion}

\subsection{A. Tchekhovskoy -- What Sets the Power of Jets from Accreting Black 
Holes? (invited)}

It is crucial to understand what sets the maximum power of jets, if jets are 
powered by black holes (BHs) or inner regions of the accretion disks. Jet power 
depends on magnetic field topology: dipolar geometry gives powerful jets, 
quadripolar or toroidal gives weak or no jets. Jet power increases with 
increasing BH magnetic flux. BH and a large magnetic flux give a 
magnetically-arrested accretion (MAD): the BH is saturated with flux, and the 
$B$-field is as strong as gravity. Radio-loud AGN have MADs with BH spins ($a$) 
near to 1 and radio-quiet AGN shows MADs with $a < 0.1$. Retrograde BHs appear 
to have less powerful jets while thicker disks show more powerful jets.

References: \cite{tch10}, \cite{tch11}, \cite{mck12}, \cite{tch12}.

\begin{figure}
\begin{center}
\includegraphics[width=3in]{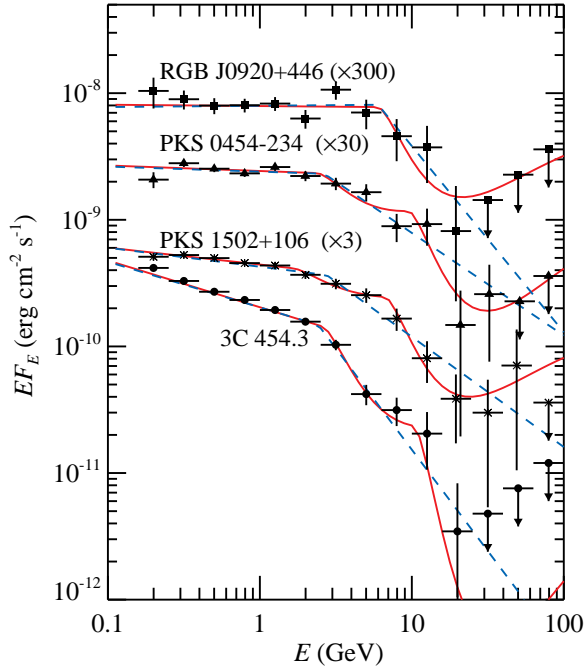}
  \caption{\Fermi-LAT SEDs of four blazars with best-fit broken power law (blue 
dash lines) and power law with absorber model (red lines). Figure 
taken from \cite{pou10}.}
    \label{fig:poutanen}
\end{center}
\end{figure}

\subsection{J. Poutanen -- \Fermi\ observations of blazars: implications for 
gamma-ray production (contribution)}

The presence in the blazar spectra of GeV breaks (see Fig.~\ref{fig:poutanen} 
for examples) produced by photon-photon absorption on BLR photons proves that 
gamma-rays are produced within the BLR. 
These spectral breaks are relatively stable 
during flares. Most GeV breaks cannot be produced by Ly-$\alpha$ photons, but 
can be produced by Lyman continuum of ionized helium. The gamma-ray emitting 
region is located at the boundary of the He$^{++}$ zone and moves away from the 
BLR high-ionization zone (moving region model). This model is consistent with 
the arrival of photons with energy larger than 10 GeV in the end of the flare. 
Alternative interpretations that gamma-rays are produced together or even 
after radio flares are ambiguous.

References: \cite{abd10b}, \cite{abd11}, \cite{pou10}, \cite{ste11}.

\subsection{S. Trippe -- Probing the mm/radio polarization of active galactic 
nuclei (contribution)}

The Plateau de Bure Interferometer (PdBI) is located in the south-east of 
France, consists of six 15-m antennas, and observes with dual-polarization 
single-band receivers in the frequency range 80 – 371 GHz, with a maximum 
baseline of 760 m ($\sim 0.2'' - 7''$ angular resolution). Each PdBI observation 
collects polarization information on calibration quasars (since 2007), thus 
resulting in a large mm polarization survey at 1.3 / 2 / 3 mm ``for free.'' Four 
years of polarization monitoring data of AGN were used for this study. 
Polarization is detected in almost all sources (73 out of 86 quasars), 
consisting of 316 detections out of 441 source measurements. Fast variability in 
flux and polarization are observed with almost identical fluctuation rates, 
implying similar spatial scales are probed. Polarization variations indicate 
strong shocks and allow for estimates of the shock parameters. Very large 
rotation measures (up to 400,000 radians m$^{-2}$) are measured and allow an 
estimate of the outflow geometries (spherical/conical outflows).

References: \cite{tri10}, \cite{tri12}.

\subsection{M. Su -- Fermi Gamma-Ray bubbles, jets, and lines in the Milky Way 
(contribution) \label{sec:su}}

The Fermi bubbles are giant gamma-ray structures with sharp edges discovered 
using data from the \Fermi-LAT. They rise up \& down from the Galactic center 
with extents of $\sim50$ degrees ($\sim$8.5 kpc), are well centered on longitude 
zero and close to latitude zero, and imply the acceleration of TeV electron 
energy particles. They could be related to a jet or outflow activity from the 
Galactic center. The gamma-ray bubbles have counterparts at microwave 
frequencies (the WMAP haze, confirmed by \Planck). Sharp edges are observed in 
X-ray also utilizing ROSAT data, and several small \textit{XMM-Newton} pointings 
have recently been obtained in order to study this in more detail. The basic 
question surrounding the origin of the bubbles is whether they are jet or 
wind/outflow driven. A gamma-ray jet / cocoon feature was recently detected in 
the \Fermi-LAT data offering additional clues. Overall, the Fermi bubbles are 
analogous to large-scale structures observed in more distant AGN, e.g., Cen~A, 
and connections are being sought. 

References: \cite{su10}, \cite{su12}.

\begin{acknowledgments}
We would like to acknowledge the support from the IAU, the JD6 participants 
for their contributions, and members of the organizing committee: 
Ed Fomalont, Luigi Foschini, Marcello Giroletti, Xiaoyu Hong (co-chair), Seiji 
Kameno, Matthias Kadler, Yuri Kovalev, Laura Maraschi (co-chair), David Paneque, 
Maria Rioja, Eduardo Ros, {\L}ukasz Stawarz, Meg Urry, and Anton Zensus.
T.C. was supported at NRL by NASA DPR S-15633-Y.
\end{acknowledgments}


\begin{thebibliography}{}

\bibitem[Abdo et al. (2009a)]{abd09a} Abdo, A.~A., Ackermann, M., Ajello, M., et 
al.\ 2009a, \apj, 699, 976 (PMN J0948+0022)

\bibitem[Abdo et al. (2009b)]{abd09b} Abdo, A.~A., Ackermann, M., Ajello, M., et
al.\ 2009b, \apjl, 707, L142 (LAT radio-loud NLS1s)

\bibitem[Abdo et al. (2010a)]{abd10a} Abdo, A.~A., Ackermann, M., Ajello, M., et
al.\ 2010a, {\it Science}, 328, 725 (Cen~A lobes)

\bibitem[Abdo et al. (2010b)]{abd10b} Abdo, A.~A., Ackermann, M., Ajello, M., et 
al.\ 2010b, \apj, 710, 1271 (LAT blazar spectra)

\bibitem[Abdo et al. (2011)]{abd11} Abdo, A.~A., Ackermann, M., Ajello, M., et 
al.\ 2011, \apjl, 733, L26 (3C~454.3 flare)

\bibitem[Abraham \& Romero (1999)]{abr99} Abraham, Z., \& Romero, G.~E.\ 1999, 
\aap, 344, 61

\bibitem[Abramowski et al. (2012)]{abr12} Abramowski, A., Acero, F., Aharonian, 
F., et al.\ 2012, \apj, 746, 151 (M87)

\bibitem[Acciari et al. (2011)]{acc11} Acciari, V.~A., Aliu, E., Arlen, T., et 
al.\ 2011, \apj, 738, 25 (Mrk 421)

\bibitem[Ackermann et al. (2011a)]{ack11a} Ackermann, M., Ajello, M., Allafort, 
A., et al.\ 2011a, \apj, 741, 30 (radio-gamma connection)

\bibitem[Ackermann et al. (2011b)]{ack11b} Ackermann, M., Ajello, M., Allafort, 
A., et al.\ 2011b, \apj, 743, 171 (2LAC)

\bibitem[Ackermann et al. (2012)]{ack12} Ackermann, M., Ajello, M., Allafort, A.,
et al.\ 2012, \apj, 747, 104 (LAT Seyferts)

\bibitem[Aleksi{\'c} et al. (2012)]{ale12} Aleksi{\'c}, J., Alvarez, E.~A., 
Antonelli, L.~A., et al.\ 2012, \aap, 542, A100 (Mrk 421)

\bibitem[Cai et al. (2007)]{cai07} Cai, H.-B., Shen, Z.-Q., Sudou, H., et al.\ 
2007, \aap, 468, 963

\bibitem[Cerutti et al. (2012)]{cer12} Cerutti, B., Werner, G.~R., Uzdensky,
D.~A., \& Begelman, M.~C.\ 2012, \apjl, 754, L33

\bibitem[D'Ammando et al. (2012)]{dam12} D'Ammando, F., Orienti, M., Finke, J., 
et al.\ 2012, \mnras, 426, 317

\bibitem[Dermer et al. (2012)]{der12} Dermer, C.~D., Murase, K., \& Takami, H.\
2012, \apj, 755, 147

\bibitem[Ghirlanda et al. (2011)]{ghi11} Ghirlanda, G., Ghisellini, G.,
Tavecchio, F., Foschini, L., \& Bonnoli, G.\ 2011, \mnras, 413, 852

\bibitem[Ghisellini \& Tavecchio (2008)]{ghi08} Ghisellini, G., \& Tavecchio, F.\
2008, \mnras, 386, L28

\bibitem[Ghisellini et al. (2009)]{ghi09} Ghisellini, G., Tavecchio, F., Bodo,
G., \& Celotti, A.\ 2009, \mnras, 393, L16

\bibitem[Giannios et al. (2009)]{gia09} Giannios, D., Uzdensky, D.~A., \&
Begelman, M.~C.\ 2009, \mnras, 395, L29

\bibitem[Giannios et al. (2010)]{gia10} Giannios, D., Uzdensky, D.~A., \&
Begelman, M.~C.\ 2010, \mnras, 402, 1649

\bibitem[Hayashida et al. (2012)]{hay12} Hayashida, M., Madejski, G.~M., 
Nalewajko, K., et al.\ 2012, \apj, 754, 114

\bibitem[HESS Collaboration (2012)]{hess12} HESS Collaboration: Abramowski, A.,
Acero, F., et al.\ 2012, \aap, 545, A103

\bibitem[Jorstad et al. (2011)]{jor11} Jorstad, S., Marscher, A., Agudo, I., 
\& Harrison, B.\ 2011, 2011 Fermi Symposium, eConf C110509, arXiv:1111.0110

\bibitem[Kardashev et al. (2013)]{kar13} Kardashev, N.~S., Khartov, V.~V., 
Abramov, V.~V., et al.\ 2013, {\it Astronomy Reports}, 57, 153

\bibitem[Karouzos et al. (2011)]{kar11} Karouzos, M., Britzen, S., Witzel, A., 
Zensus, J.~A., \& Eckart, A.\ 2011, \aap, 529, A16

\bibitem[Kovalev et al. (2009)]{kov09} Kovalev, Y.~Y., Aller, H.~D., Aller, 
M.~F., et al.\ 2009, \apjl, 696, L17

\bibitem[Le{\'o}n-Tavares et al. (2011)]{leo11} Le{\'o}n-Tavares, J., Valtaoja, 
E., Tornikoski, M., L{\"a}hteenm{\"a}ki, A., \& Nieppola, E.\ 2011, \aap, 532, 
A146

\bibitem[Lister et al. (2009)]{lis09} Lister, M.~L., Cohen, M.~H., Homan, D.~C., 
et al.\ 2009, \aj, 138, 1874

\bibitem[Lister et al. (2011)]{lis11} Lister, M.~L., Aller, M., Aller, H., et 
al.\ 2011, \apj, 742, 27

\bibitem[Liu et al. (2008)]{liu08} Liu, H.~T., Bai, J.~M., \& Ma, L.\ 2008, \apj,
688, 148

\bibitem[Mahony et al. (2010)]{mah10} Mahony, E.~K., Sadler, E.~M., Murphy, T., 
et al.\ 2010, \apj, 718, 587

\bibitem[Max-Moerbeck (2012)]{max12} Max-Moerbeck, W.\ 2012, {\it BAAS}, 219, 
\#321.05

\bibitem[McConnell et al. (2012)]{mcc12} McConnell, D., Sadler, E.~M., Murphy, 
T., \& Ekers, R.~D.\ 2012, \mnras, 422, 1527

\bibitem[McConville et al.(2011)]{mcc11} McConville, W., Ostorero, L., Moderski, 
R., et al.\ 2011, \apj, 738, 148

\bibitem[McKinney et al. (2012)]{mck12} McKinney, J.~C., Tchekhovskoy, A., \& 
Blandford, R.~D.\ 2012, \mnras, 423, 3083

\bibitem[Nalewajko et al. (2012)]{nal12} Nalewajko, K., Begelman, M.~C., Cerutti,
B., Uzdensky, D.~A., \& Sikora, M.\ 2012, \mnras, 425, 2519

\bibitem[Nieppola et al. (2011)]{nie11} Nieppola, E., Tornikoski, M., Valtaoja, 
E., et al.\ 2011, \aap, 535, A69

\bibitem[Orienti et al. (2013)]{ori13} Orienti, M., Koyama, S., D'Ammando, F., 
et al.\ 2013, \mnras, 428, 2418

\bibitem[Paneque et al. (2012)]{pan12} Paneque, D., Ballet, J., Burnett, T., et 
al.\ 2012, 4th Fermi Symposium, eConf C121028, arXiv:1304.4153

\bibitem[Pavlidou et al. (2012)]{pav12} Pavlidou, V., Richards, J.~L., 
Max-Moerbeck, W., et al.\ 2012, \apj, 751, 149

\bibitem[Poutanen \& Stern (2010)]{pou10} Poutanen, J., \& Stern, B.\ 2010, 
\apjl, 717, L118

\bibitem[Reimer (2007)]{rei07} Reimer, A.\ 2007, \apj, 665, 1023

\bibitem[Sudou \& Iguchi (2011)]{sud11} Sudou, H., \& Iguchi, S.\ 2011, \aj, 
142, 49

\bibitem[Stern \& Poutanen (2011)]{ste11} Stern, B.~E., \& Poutanen, J.\ 2011, 
\mnras, 417, L11

\bibitem[Su et al. (2010)]{su10} Su, M., Slatyer, T.~R., \& Finkbeiner, D.~P.\ 
2010, \apj, 724, 1044

\bibitem[Su \& Finkbeiner (2012)]{su12} Su, M., \& Finkbeiner, D.~P.\ 2012, 
\apj, 753, 61

\bibitem[Takeuchi et al. (2012)]{tak12} Takeuchi, Y., Kataoka, J., Stawarz,
{\L}., et al.\ 2012, \apj, 749, 66

\bibitem[Tavecchio \& Mazin (2009)]{tav09} Tavecchio, F., \& Mazin, D.\ 2009,
\mnras, 392, L40

\bibitem[Tchekhovskoy et al. (2010)]{tch10} Tchekhovskoy, A., Narayan, R., \& 
McKinney, J.~C.\ 2010, \apj, 711, 50

\bibitem[Tchekhovskoy et al. (2011)]{tch11} Tchekhovskoy, A., Narayan, R., \& 
McKinney, J.~C.\ 2011, \mnras, 418, L79

\bibitem[Tchekhovskoy \& McKinney (2012)]{tch12} Tchekhovskoy, A., \& McKinney, 
J.~C.\ 2012, \mnras, 423, L55

\bibitem[Trippe et al. (2010)]{tri10} Trippe, S., Neri, R., Krips, M., et al.\ 
2010, \aap, 515, A40

\bibitem[Trippe et al. (2012)]{tri12} Trippe, S., Neri, R., Krips, M., et al.\ 
2012, \aap, 540, A74

\bibitem[Zhao et al. (2011)]{zha11} Zhao, G.-Y., Chen, Y.-J., Shen, Z.-Q., et 
al.\ 2011, {\it Journal of Astrophysics \& Astronomy}, 32, 61

\bibitem[Zhao et al. (2013)]{zha13} Zhao, G.-Y., Chen, Y.-J., Shen, Z.-Q., et 
al.\ 2013, IAU Symposium, 290, 367

\end{thebibliography}
\end{document}